\documentclass[12pt,preprint]{aastex}
\shorttitle{BATSE $z>$8 GRBs?}
\shortauthors{Ashcraft \& Schaefer}

\begin{document}

%% LaTeX will automatically break titles if they run longer than
%% one line. However, you may use \\ to force a line break if
%% you desire.
\title{Are There Any Redshift $>$8 Gamma-Ray Bursts in the BATSE Catalog?}

%% Use \author, \affil, and the \and command to format
%% author and affiliation information.
%% Note that \email has replaced the old \authoremail command
%% from AASTeX v4.0. You can use \email to mark an email address
%% anywhere in the paper, not just in the front matter.
%% As in the title, you can use \\ to force line breaks.
\author{Teresa Ashcraft}
\author{Bradley E. Schaefer}
\affil{Physics and Astronomy, Louisiana State University,
    Baton Rouge, LA, 70803}

\begin{abstract}

Several luminosity indicators have been found for Gamma-Ray Bursts (GRBs) wherein measurable light curve and spectral properties are well-correlated with the peak luminosity.  Several papers have each applied one different luminosity relation to find redshifts for BATSE GRBs and claim to identify specific bursts with $z > 8$. The existence of such high redshift events is not surprising, as BATSE has enough sensitivity to see them and GRBs are expected out to the redshift of the first star formation. To improve results we used five luminosity relations with updated calibrations to determine redshifts with error bars.  Combining these relations, we calculated the redshifts of 36 BATSE GRBs with claimed $z > 8$.  Our results include 13 bursts with our derived best redshift $z_{best }> 8$, which looks promising at first.  But the calculated redshift uncertainties are significantly large in these selected cases.  With only one exception, all of our bursts have $ z_{1\sigma low} \leq 9$.  The one exception (BATSE trigger  2035) is likely a short duration burst at $z \gtrsim 4$.  Our best case for a very high redshift event is BATSE trigger 3142 with $z_{best} \geq 20$ and $z_{1\sigma low}=8.9$, however we can only say $z>4.1$ at the two-sigma confidence level.  In all, we cannot point toward any one BATSE burst as confidently having $z > 8$.  One implication is to greatly weaken prior claims that GRBs have a steeply rising rate-density out to high redshifts.

\end{abstract}

\section{Introduction}

	Long duration Gamma-Ray Bursts (GRBs) are expected to be visible to high redshifts.  This is because they are caused by the core collapses of massive stars, which have very short lifetimes (Woosley \& Bloom 2006), so the first bursts in our Universe should be shortly after the first star formation.  The {\it WMAP} detection of reionization at redshifts $20^{+10}_{-9}$ (Kogut et al. 2003), $17 \pm 4$ (Spergel et al. 2003), or $\sim 10-20$ (Spergel et al. 2006) set the scale for the redshift at which the first significant star formation occurs.  On the reasonable assumption that the GRB rate is proportional to the massive star formation rate, models of star formation in our Universe can be translated into predictions as to the relative rate of GRBs as a function of redshift.  Detailed calculations show that the first GRBs should have redshifts of $ > 20$ (Bromm \& Loeb 2002; 2006) or $\sim 60$ (Naoz \& Bromberg 2007).  GRBs have already been identified at $z=6.295$ (Totani et al. 2006) and $z=6.6$ (Grazian et al. 2006; Tanvir et al. 2006).  The lack of bursts with higher spectroscopic redshifts is likely due to the severe observational difficulties in recognizing the events and measuring their spectra.  Detailed models for the fraction of Swift bursts with $z>8$ give 10\% (Bromm \& Loeb 2002), ~3\% (Le \& Dermer 2007), and $\sim$5\% (Bromm \& Loeb 2006).
	
	The BATSE detectors onboard the {\it Compton Gamma Ray Observatory} have a deep sensitivity threshold and most of the sky was monitored for nine years, so there are likely to be many high redshift GRBs hidden in its catalog.  BATSE can detect typical GRBs out to redshifts of 20-30 (Lamb \& Reichart 2000). The fraction of BATSE bursts with $z>8$ should be 4\% (Bromm \& Loeb 2002).  With 1637 bursts in the fourth BATSE catalog (Paciesas et al. 1999), we expect $\sim 70$ $z>8$ GRBs.  
	
	One way to recognize high redshift BATSE bursts is to use the various luminosity indicators.  Just as with Cepheids and supernovae, readily observable light curve and spectral properties are well-correlated with the luminosity, so that the observed brightness plus the inverse-square law will allow a luminosity distance to be derived.  For some fiducial cosmology with a distance/redshift relation, the redshift can then be calculated.  The luminosity indicators include the spectral lag time (Norris, Marani, \& Bonnell 2000; Norris 2002; Schaefer 2004), the variability (spikiness) in the light curve (Fenimore \& Ramirez-Ruiz 2000; Reichart et al. 2001; M\'{e}sz\'{a}ros et al. 2002; Kobayashi, Ryde, \& MacFadyen 2002; Li \& Paczynski 2006), the photon energy of the spectral peak $E_{peak}$ (Schaefer 2003; 2007; Yonetoku et al. 2004), the minimum rise time in the light curve (Schaefer 2007), and the number of peaks in the light curve (Schaefer 2007).  A very-high-luminosity burst is recognizable as having a near-zero lag, a very spiky light curve, a high $E_{peak}$, a very fast rise time, and many peaks.  The luminosity relations (i.e., luminosity as a function of the luminosity indicator) are calibrated with 69 GRBs that have spectroscopic redshifts (Schaefer 2007).  The luminosity relations vary substantially in their accuracy, and one of them (the number of peaks) only provides a lower limit on the luminosity.  
	
	Various previous researchers have applied one of the luminosity indicators to a large number of BATSE bursts (Fenimore \& Ramirez-Ruiz 2000; Schaefer, Deng, \& Band 2001; Lloyd-Ronning, Fryer,, \& Ramirez-Ruiz 2002; Band, Norris, \& Bonnell 2004; Yonetoku et al. 2004; Kocevski \& Liang 2006).  The goal was to measure the redshifts of BATSE bursts for various demographic reasons.  One such purpose was to measure the burst event rate as a function of redshift so as to derive the evolution of the universal star formation rate, while another purpose was to determine the burst luminosity function and its evolution.  These studies all identified individual bursts that apparently had very high redshifts.
	
	Band, Norris, \& Bonnell (2004) used the lag-luminosity relation to derive the redshifts to 1194 BATSE bursts, for which they identified 44 bursts with $z>8$ (4\%) and 17 bursts with $z>12$ (1.4\%).  Yonetoku et al. (2004) used the $E_{peak}$-luminosity relation to derive the redshifts of 745 BATSE bursts, for which 77 have $z>8$ (10\%) while none have $z>12$.  Fenimore \& Ramirez-Ruiz (2000) used the variability-luminosity relation to derive the redshifts of 220 BATSE bursts, finding 38 bursts with $z>7$ (17\%) and 19 bursts with $z \geq12$ (8.6\%).  If correct, these very high redshifts would be of high interest and useful for cosmology.
	
	Unfortunately, as these researchers were well aware at the time, the highest derived redshifts have substantial uncertainties.  To give an example, consider the lag-luminosity relation, where the burst luminosity is inversely proportional to the lag.  The highest luminosity events will all have small lags, but the measurement uncertainty of lags is roughly a constant (depending on the number of detected photons and the light curve shape) so it will be impossible to distinguish between a very high luminosity burst (with a very short lag) at a very high redshift and a moderately high luminosity burst (with a moderately short lag) at a moderately high redshift.  To illustrate this, consider BATSE  trigger 2850 for which the lag was measured to be 0.008 seconds, but for which the one-sigma range is between -0.044 and 0.092 seconds (Band, Norris, Bonnell 2004).  The best fit value for the lag happened to be near zero out of a large uncertainty range and this was followed through to yield a formal best fit redshift of 22.1.  But the one-sigma range of lag runs to 0.092 sec, and this corresponds to a luminosity 11.5 times fainter and a redshift of $\sim 0.1$.  Thus, in the ordinary course of Gaussian errors, many moderate redshift bursts with a lag of, say, 0.1 sec will appear at very high redshift due to measured lags of 0.01 sec or even 0.001 sec.  As such, normal measurement errors will smear the burst redshift distribution so as to artificially populate the highest redshifts.  Thus, keeping track of the uncertainties is important.  Only the Yonetoku et al. (2004) paper actually quoted their redshift uncertainties. Of their 77 bursts with the best estimate of the redshift as $z>8$, only 9 have the one-sigma range above $z=8$, while only one has its two-sigma range barely requiring $z>8$.
	
	Another challenge for the idea of these BATSE bursts being at very high redshift is that derived redshifts from multiple luminosity indicators should agree that the redshift is high.  For example, it would not be convincing to have a redshift of $12 \pm 1$ derived from the burst's lag if the variability and $E_{peak}$ values are pointing to $z \approx 3$.  However, if all the luminosity indicators unanimously pointed to $z \approx 12$, then we would have some confidence that the redshift is indeed very high.  To date, no such cross-comparison has been reported.  The use of multiple indicators has a further advantage that the independent measures of the luminosity can be combined as a weighted average so as to substantially reduce the size of the derived one-sigma region.
	
	Another problem with the old reported BATSE redshifts is that they were made with old definitions and calibrations.  Many more GRBs with spectroscopic redshifts are now available for calibrating the luminosity relations, and realistic systematic errors are now known (Schaefer 2007).
	
	This paper reports on our concerted attempt to identify very high redshift GRBs in the BATSE database.  In particular, we tested whether any of the specific bursts reported to have very high redshifts can indeed be confidently concluded to be at very high redshift.   For this, we are using five luminosity indicators so as to look for consistency.  We are using the latest definitions and calibrations for the luminosity relations.  We are keeping track of the uncertainties in each indicator so as to get a best fit redshift as well as a quantified one-sigma range.  The result is that we can confidently test whether the previously identified very-high-redshift BATSE events are really at such high redshifts.
	
\section{GRB Selection and Properties}

	Initially, we looked through the published lists of BATSE burst redshifts for cases where one burst was consistently quoted to be at very high redshifts.  No such cases were found.  This in itself is a foreshadowing of our final answer.  Nevertheless, we soon found a variety of the usual problems concerning consistencies between measurements and old calibrations.  Thus, we realized that we would have to derive many of the light curve properties on our own and then apply our recent calibrations.  

	We decided to look at the 36 BATSE bursts with the highest chances of being at very high redshifts as based on prior publications (Band et al. 2004; Yonetoku et al. 2004; Fenimore \& Ramirez-Ruiz 2000).  These three prior publications quote independent redshifts for hundreds of BATSE bursts based on one of three GRB luminosity indicators (lag, $E_{peak}$, and variability respectively).  We selected the 12 GRBs with the highest redshifts from each of the three sources.  (Again, the lack of overlap in these selections point to a lack of confident high-z bursts.)  Two bursts originally chosen did not have any available $E_{peak}$, so the next two highest redshift bursts from the same source were selected in their place.   The 36 selected bursts all have $z > 8$ as reported by one of the three papers.  This selection criterion will inevitably lead to bursts with one extreme indicator, discrepancies between indicators, and large error bars in redshift. 
	
	Table 1 contains the measurable burst properties leading up to the derived bolometric peak flux. Column 1 contains the BATSE trigger number that identifies each burst.  A footnote identifies which of the papers was used for selecting the burst.  Column 2 lists the $T_{90}$ of each burst, with this duration measuring the interval in seconds between the time when 5\% to 95\% of the total counts have been detected.  Column 3 has the observed peak flux over a one second time interval in the energy range of 50-300 keV ($P_{50-300}$ with units of photon cm$^{-2}$ s$^{-1}$).   Both $P_{50-300}$ and $T_{90}$ have been taken from the Fourth BATSE catalog and the 'current catalog'\footnote{http://cossc.gsfc.nasa.gov/docs/cgro/batse/BATSE\_Ctlg/index.html} (Paciesas et al. 1999).  GRB spectra can be reasonably described as smoothly broken power laws, as given by the ÔBand modelÕ (Band et al. 1993). In this model, the low energy power law index is $\alpha$, the high energy power law index is $\beta$, and the photon energy of peak spectral flux is $E_{peak}$. Columns 3-5 contain the values for $E_{peak}$, $\alpha$, and $\beta$.  These values are those reported by Band, Norris, \& Bonnell (2004)\footnote{See also http://cossc.gsfc.nasa.gov/docs/cgro/analysis/lags/}.  Of these, the majority are from unpublished measures by Robert Mallozzi (c.f. Mallozzi 1995).  The values placed in square brackets are typical default values adopted when the value itself was not measured.  The spectral parameters and the peak fluxes were used to find the bolometric peak fluxes ($P_{bolo}$) as in equation 6 of Schaefer (2007).  The value was calculated for the best redshift (see Table 4), although the dependancy on the assumed redshift is relatively small. The resulting $P_{bolo}$ (in column 7) is over a one second time interval and has units of erg cm$^{-2}$ s$^{-1}$.  The uncertainty of $P_{bolo}$ is found by propagating the uncertainties of $P_{50-300}$, $E_{peak}$, $\alpha$, and $\beta$. 
	
	Table 2 lists the luminosity indicators for each burst. ($E_{peak}$ is also a luminosity indicator, although it is tabulated in Table 1.) Column 2 contains the lag time ($\tau_{lag}$) of a GRB which is the time in seconds between the hard (100-300 keV ) and soft photon (25-50 keV) light curves (Norris, Marani, \& Bonnell 2000).  The values listed are taken from Band et al. (2004), with the quoted error bars being for the positive lag direction.  These values agree within error bars with the lags reported for these same bursts by Hakkila et al. (2007). If Band et al. (2004) found a negative or zero lag, we took the lag to be 0.001 seconds. Column 3 has the variability ($V$) which is dimensionless and measures the spikiness of the burst light curve. It is measured by calculating the variance of the light curve from a smooth version the light curve (Fenimore \& Ramirez-Ruiz 2000).  We used the BATSE 25-300 keV light curves with the definition of variability as given in equation 9 of Schaefer (2007).  Column 4 contains the minimum rise time ($\tau _{RTmin}$) which is the shortest amount of time in seconds for the light curve to rise by half its peak.  The number of significant peaks ($N_{peak}$) in each burst light curve is listed in column 5.  A peak is counted if its peak flux lies above the background is at least 25\% of the largest peaks flux and there is a dip between it and the surrounding peaks of at least 25\% of the peak flux. We have calculated $V$, $\tau _{RTmin}$, and $N_{peak}$ from 0.064 ms concatenated BATSE light curves for 25-300 keV\footnote{Available at http://cossc.gsfc.nasa.gov/docs/cgro/batse/batseburst/sixtyfour\_ms/index.html.}.
	
\section{Luminosity Relations}

	The luminosity relations are calibrated equations that give the burst's peak luminosity ($L$) as a function of the measured luminosity indicators.  To date, eight different luminosity relations have been proposed and verified.  In this paper, we will be using only five of these relations, the same five as used by Schaefer (2007) for constructing the 69 GRB Hubble diagram.  The $E_{peak}-E_{\gamma}$ relation of Ghirlanda, Ghisellini, \& Lazzati (2004) was not used as the required jet breaks in the afterglow light curves were not seen for BATSE bursts.  The $E_{peak}-E_{iso,\gamma}$ relation of Amati (Amati et al. 2002; Amati 2006) was not used as its luminosity indicator is duplicated in the $E_{peak}-L$ relation which we are already using.  The $E_{peak}-T_{0.45}-L$ relation of Firmani et al. (2006) is not used as its input and physics is similar to that of the $E_{peak}-L$ relation which we are already using.
	
	The five luminosity relations that we use are summarized in Table 3.  The lag-luminosity relation is fairly tight, but the bunching at near-zero lags makes for difficulties in distinguishing high-luminosity and very-high-luminosity bursts.  The variability-luminosity relation is one of the poorest of the eight in terms of accuracy, but there is a substantial spread in variability between the high-luminosity and the very-high-luminosity bursts.  The $E_{peak}$-luminosity relation was originally predicted theoretically and then observationally confirmed (Schaefer 2003), with later verification (Yonetoku et al. 2004).  The $\tau_{RTmin}$-luminosity relation was also predicted and confirmed in 2002, although this relation is also relatively poor in accuracy.  The $N_{peak}$-luminosity relation is another prediction from 2002, and even though it only  provides a {\it limit} on the luminosity, the relation works in the right direction for the purposes of this paper as a many-peaked bursts must have very high luminosity.
		
	The luminosity relations must be calibrated with bursts of known redshift (and hence with known peak luminosity).  Schaefer (2007) has collected 69 GRBs for use in calibrating the five luminosity relations, and we have adopted these calibrations (see Table 3).  Imporantly, these calibrations are for the same definitions of luminosity indicators as we use in this paper.  So, for example, we should not use the $V$ values given by Fenimore \& Ramirez-Ruiz in the equation quoted in Table 3. 
	
	A possible concern is that the luminosity relations are calibrated with a variety of different satellites and then applied to BATSE.  For light curve luminosity indicators, the only systematic difference could arise from the use of different energy bands.  But here, the chosen bands are closely similar (see Table 5 of Schaefer 2007), so systematic problems should be negligible.  For the $E_{peak}$ relations, the different satellites are all measuring the same quantity without any known biases, so as long as the (possibly asymmetric) error bars are propagated, the joint calibrations should apply to any individual satelliteÕs data.  The real proof that the calibrations are not satellite dependent is that Schaefer (2007) found the best fit calibrations for each satellite to be identical to within the error bars.

	The accuracy of the derived redshifts from combining all five luminosity relations is substantially better than can be obtained by any one relation alone (Schaefer 2007).  The overall accuracy can be quantified by comparing derived redshifts versus the spectroscopic redshifts.  For 69 GRBs, the one-sigma scatter about the known redshifts is 26\% (Schaefer 2007).  And the reduced chi-square of this comparison is close to unity, with the implication that the calculated error bars are realistic.  This is the basis for knowing that the luminosity relations are reliable and accurate to within the stated error bars.
	
	The luminosity relations do not apply to short duration bursts, presumably because they have a different physical mechanism.  The dividing line between long-duration and short-duration bursts is around 2 seconds as measured in the BATSE rest frame (Kouveliotou et al. 1993).  Presumably, such a division is better done in the rest frame of the burst, which equals $2/(1+z)$ s, where $z$ is the typical redshift of BATSE bursts.  For the typical redshifts of BATSE bursts of roughly 1, the division between long and short should be roughly 1 second in the burst rest frame.
				
\section{Redshifts}

	Each of the five luminosity relations will yield a luminosity with its error bar.  For each of these luminosities (along with $P_{bolo}$) we can derive a luminosity distance as $(L/[4\pi P_{bolo}])^{\onehalf}$.  Then, we can get a distance modulus and its uncertainty ($\mu \pm \sigma_{\mu}$) for each of the 36 GRBs.  But each of the luminosity indicators must be corrected to the rest frame of the burst.  So we have set this up by evaluating $\mu$ and $\sigma_{\mu}$ for all redshifts (each corresponding to a distance modulus $\mu_z$) from 0.01 to 20 at an interval of 0.01.  For each luminosity relation and each trial redshift, we can evaluate the chi-square as $[(\mu - \mu_z)/\sigma_{\mu})]^2$.  The resultant chi-squares for all five luminosity relations are then coadded to get a total chi-square which is a function of redshift.  Our best redshift ($z_{best}$) is the value of $z$ for which the chi-square is the smallest.  Our one-sigma confidence region for the derived redshift is the range of redshift over which the chi-square is within 1.0 of the minimum chi-square.  Our two-sigma region is where the chi-square is within 4.0 of the minimum.  For each individual luminosity relation, we can similarly derive a best redshift ($z_{lag}$, $z_{V}$, $z_{Epeak}$, $z_{RT}$, and $z_{Npeak}$) and the one-sigma confidence interval.  These are listed in Table 4.  The error bars are generally asymmetric, so we have represented the one-sigma confidence interval in parentheses immediately after the optimal value.  Some of these redshifts or their confidence ranges are open-ended to high redshift, and we represent this by capping the upper limit at a redshift of 20.
	  
	  How consistent are the five derived redshifts from the five luminosity relations?  A quantitative way to answer this question is to look at the chi-squares for $z_{best}$.  With three degrees of freedom for our chi-square fit (the limit from $N_{peak}$ is never constraining), we expect that the average chi-square will be around 3, while a three-sigma discrepancy would require the chi-square to be greater than 3+9=12.  For our 36 bursts, we find that the typical chi-square is around 3 and the largest chi-square values (for BATSE triggers 2345 and 7390) are near 7.  From this, we conclude that all of our luminosity indicators are giving consistent redshifts.
	  
	The one-sigma uncertainty ranges on $z_{best}$ can be fairly large.  One reason for this is that the very-high luminosity case gives near-zero indicators for both lag and risetime, so that the usual measurement uncertainties only put a lower limit on the luminosity.  Also, the high $E_{peak}$ values are difficult to measure with accuracy.  Another reason is that the slope of the redshift versus luminosity distance relation changes at high redshifts, so that the upward error bars in redshift extend over a larger numerical range.  As such, our expected large redshift uncertainties for the 36 bursts in this paper are {\it not} characteristic of normal bursts.
	  
	  In 13 cases, the $z_{best}$ value is $>8$, which would be of interest, but the one-sigma range extends down to moderate redshift.  For example, BATSE trigger 1734 has a best redshift of 14.5 but a one-sigma confidence region of $>8.3$ and a two-sigma confidence region of $>4.1$.  In such a case, we cannot confidently identify the burst as $z>8$, because the redshift could be smaller at only the 1.1-sigma confidence level.  That is, the redshift of trigger 1734 can easily be 4- 6, which is no longer novel.  For convenience, we have added a last column to Table 4 which lists the redshifts of the lower edges of the one-sigma confidence regions ($z_{1\sigma low}$).  Let us examine the redshifts for four cases so as to see the range of our results.  
	  
	  The first case (BATSE trigger 111) was originally selected because its $E_{peak}$ value was sufficiently high for Yonetoku et al. (2004) to quote that $z_{Epeak}=11.2$, whereas we get a more moderate $z_{Epeak}=5.5$.  This difference is not due to the $E_{peak}$ value, as that from Band et al. (2004) is close to that used by Yonetoku et al.  Rather, the difference is mainly caused by the difference in the slope of the $E_{peak}-L$ relation, where they used $2.0 \pm 0.2$ while we used $1.68 \pm 0.05$.  (The two slopes for the relation are different because Yonetoku et al. [2004] only used 11 GRBs for calibration, whereas Schaefer [2007] used 64 GRBs for calibration.)  In any case, Yonetoku et al. gave the lower edge of their one-sigma range for $z_{Epeak}$ as 4.72, so there is no significant discrepancy.  In the meantime, all the other four luminosity relations point to a moderately low redshift, as the burst has a long lag, smooth light curve, slow rise time, and only one peak.  The $z_{best}$ is 1.4, which is only just outside the one-sigma range for $z_{Epeak}$. In this case, $z_{Epeak}$ has a large deviation from $z_{best}$, and this is not surprising because the burst was selected for its extreme $z_{Epeak}$.
	  
	  The second case (BATSE trigger 493) was originally selected because Yonetoku et al. reported that $z_{Epeak}=11.9$, even though the lower edge of their one-sigma range was to 4.6.  We get $z_{Epeak}=4.3$ with a somewhat smaller value for $E_{peak}$.  All five individual redshifts are consistent with $z_{best}=3.7$.  For this redshift, the duration in the burst rest frame will be $T_{90}/(1+z)=1.0$ s, which still allows the event to be a long duration GRB.  Our final redshift has a one-sigma range of 2.3-6.1, so we are confident that this GRB is not at high redshift.    
	  
	  The third example is our best case for a very high redshift GRB, BATSE trigger 3142.  Our $z_{best}$ is $>20$, which appears to be of interest.  However, $z_{1\sigma low}=8.9$ while a two-sigma limit is $>4.1$, so a reasonable idea is that the burst has an 'unexciting' redshift.  Nevertheless, the burst has a very short lag, a high variability, a high $E_{peak}$, a very fast rise time, and 8 peaks, so this remains a good candidate for a very high redshift burst.

	The fourth case is BATSE trigger 2035, which has a $z_{best} > 20$ and $z_{1\sigma low} =14.9$.  This burst has a very short lag, a very high variability, and the shortest of the rise times, all of which point to a very high luminosity GRB.  When combined with the faint $P_{bolo}$, these three indicators point to $z\gtrsim10$.  However, BATSE trigger 2035 is a burst whose duration is near the low end of the range for long-duration bursts, with $T_{90}=5.1$ s.  If the burst were indeed at $z=14.9$, then the time dilation of the light curve would imply that the duration was 0.3 s in the burst's rest frame and that the event would then be a short-duration burst.  But the luminosity relations do not apply to short bursts, so we could not then derive the high redshift.  Thus, we have a logical inconsistency with BATSE trigger 2035 being a long-duration GRB with $z\gtrsim 4$.  Our best solution is to say that the event is actually a short duration burst with $z\gtrsim4$ or so.  In any case, this event cannot be a very-high-redshift long-duration GRB.  

	If BATSE trigger 2035 is a short duration event, then its redshift must be $z\gtrsim4$ so that its $T_{90}$ duration would have been less than one second in its rest frame.  Is it plausible to have a short burst at such high redshifts?  The reason for asking this is that the first discoveries of optical afterglows of short bursts were associated with galaxies at low redshifts ($z=0.16$ for GRB050709, Fox et al. 2005; $z=0.258$ for GRB050724, Prochaska et al. 2005; and $z=0.547$ for GRB051221, Soderberg et al. 2006), and this suggested that all short GRBs are at low redshifts.  However, the lack of any host galaxies to very deep limits in five short GRB error boxes (Schaefer 2006) indicates that many of these events are at moderate or high redshifts.  Berger et al. (2007) later obtained similar limits and concluded that many short bursts are at $z>1$.  Also, from a theoretical population synthesis approach for double neutron star binary mergers, a high proportion of short GRBs are predicted to occur at $z>1$ (O'Shaughnessy, Belczynski, \& Kalogera 2007).  In all, it is reasonable to think that BATSE trigger 2035 is a short duration GRB with $z\gtrsim4$.
	
	This same duration dilemma applies to 7 of our 36 bursts.  That is, the bursts with $T_{90} \lesssim 9$ s cannot be identified as $z>8$ events because then their durations in their rest frames would be under 1 second and they would not be long duration GRBs for which the luminosity indicators apply.  A possible solution is that these are long duration GRBs with moderate redshift or short duration bursts with moderate redshift.  Only 3 of our bursts (BATSE triggers 2035, 2850, and 2944) have $T_{90}/(1+z_{1\sigma low})<1$ s.  These events could be short duration bursts at moderate redshifts.  Or they could be long duration bursts whose moderate redshifts are only somewhat smaller than $z_{1\sigma low}$.  This last possibility is not surprising, as we expect 16\% of the bursts to have their real redshift below $z_{1\sigma low}$, and the fraction might be substantially higher due to our selection of bursts for extreme-$z$ by only one indicator.

	The goal of our research was to test the claims to identify specific BATSE bursts with very high redshifts.  When we began this work, we were hoping to be able to identify GRBs with $z >8$.  Indeed, 13 out of our 36 GRBs have $z_{best}>8$.  However, our candidate very-high-redshift bursts all have too large of error bars to decisively say they are at high redshifts.  With one exception (BATSE trigger 2035, see above), all bursts had $z_{1\sigma low} \le 9$, so we cannot confidently state that any of them are of very high redshift.  Some of these candidates might be at $z>8$, or alternatively they all might be at 'unexciting' redshifts of $\lesssim 6$.  The less interesting alternative might even be preferred because moderate-redshift events are apparently more common than high-redshift events so that normal observational scatter in the derived redshifts will have the high-$z_{best}$ population dominated by moderate-$z$ bursts. 
			
\section{Discussion}

	We have not been able to find any GRBs that are confidently at $z>8$.  But we know that roughly 4\% of the BATSE catalog should be at comparably high redshifts (Bromm \& Loeb 2002).  So where are they?  Part of the answer might be simply that our $z_{best}>8$ bursts are indeed at $z>8$, although we cannot prove that because the uncertainties are too large.  The other part of the answer is likely simply that the $z>8$ GRBs are in the fainter half of the BATSE catalog.  These fainter bursts are too faint for Band et al. to have derived their lags, too faint for Fenimore \& Ramirez-Ruiz to derive their variabilities, and too faint for Yonetoku et al. to measure their $E_{peak}$.  As such, the faint half of the BATSE catalog has not even been examined for high redshifts.  It is not clear whether the majority of the very high redshift events should appear in the faint half of the BATSE catalog (as befits their greater distance) or should appear in the brighter half of the BATSE catalog (as befits the evolution of the GRB luminosity function and the greater volume at high redshift, c.f. Schaefer 2000).  In any case, any very high redshift GRBs in the faint half of the BATSE catalog will have only larger uncertainties than those in the brighter half.  The solution as to where are the high redshift BATSE bursts is not likely to depend on the various types of Malmquist biases, as Schaefer (2007) has made detailed calculations on the Malmquist and gravitational lensing biases only to find that all the biases together nearly cancel out resulting in negligible effect.  In all, we conclude that BATSE catalog could well include many $z>8$ GRBs, but that the redshift uncertainties will always be too large to allow for the confident identification of any such GRB.
	
	What does all this say about the many demographic studies which extend to $z>8$?  The primary problem here is that the normal scatter of derived redshifts will appear to move many moderate- and high- redshift GRBs out into the very-high-redshift regime.  The large number of lower-$z$ bursts will produce many more $z_{best}>8$ events than there are $z>8$ bursts that are made to appear of low or moderate redshift.  Thus, the number of apparent very-high-$z$ bursts is likely to be dominated by contamination from lower redshifts.  As such, the demographics (and indeed even the existence) of $z>8$ bursts is largely unknown.  To a lesser degree, the same problems arise for bursts from perhaps redshift  4 to 8.  A typical consequence of this is that the GRB rate density cannot be rising as steeply out to very high redshifts as concluded by Fenimore \& Ramirez-Ruiz (2000), Schaefer, Deng, \& Band (2001), Lloyd-Ronning, Fryer, \& Ramirez-Ruiz (2002), and Yonetoku et al. (2004).  Our new conclusion effectively rejects the evidence that points to GRBs demonstrating a steady rise of the Universe's star formation rate to very high redshift.
	
	Is there any hope for using the BATSE bursts plus the luminosity indicators to get good demographics to moderate or high redshifts?  Our experience is that the reliance on any one luminosity indicator is poor.  Even when combining five luminosity indicators the resultant error in redshift is 26\%, and this still makes for substantial smearing of the underlying redshift distributions.  However, at the 26\% level, good work can be done with reasonable accuracy.  Certainly, a detailed model of the smearing caused by the redshift uncertainties must be performed.  Another requirement is that large number of BATSE bursts must be analyzed so as to avoid selection effects like in this paper.  In all, we expect that a comprehensive study of BATSE bursts can pull out high redshift demographics provided the study (a) uses many of the luminosity indicators, (b) models the effects of the usual uncertainties, and (c) includes a large number of bursts without selection by the indicators.

	We thank Robert Mallozzi, now deceased, for his measures of $E_{peak}$, $\alpha$, and $\beta$.  We acknowledge NASA grant NNG06GH07G  for support of our work.

\clearpage

\begin{deluxetable}{lllllll}
\tabletypesize{\scriptsize}
\tablecaption{Burst Properties and Bolometric Peak Flux 
\label{tbl1}}
\tablewidth{0pt}
\tablehead{
\colhead{BATSE Trigger}   &
\colhead{$T_{90}$ (s)}  &
\colhead{$P_{50-300}$}   &
\colhead{$E_{peak}$ (keV)}  &
\colhead{$\alpha$}  &
\colhead{$\beta$}  &
\colhead{$P_{bolo}$ (erg cm$^{-2}$ s$^{-1}$)} 
}
\startdata
111\tablenotemark{c}	&	98.2	$\pm$	2.3	&	0.59	$\pm$	0.06	&	93	$\pm$	11	&	-0.01	$\pm$	0.87	&	-2.8	$\pm$	0.4	&	$	1.63\times 10^{-7}  \pm  3.0\times 10^{-8}	$	\\
493\tablenotemark{c}	&	4.9	$\pm$	1.6	&	0.95	$\pm$	0.06	&	125	$\pm$	17	&	-0.69	$\pm$	0.34	&	-2.5	$\pm$	0.2	&	$	3.35\times 10^{-7}  \pm  4.7\times 10^{-8}	$	\\
1145\tablenotemark{b}	&	32.2	$\pm$	4.2	&	1.18	$\pm$	0.07	&	96	$\pm$	6	&	-0.82	$\pm$	0.25	&	-3.2	$\pm$	0.6	&	$	3.52\times 10^{-7}  \pm  4.7\times 10^{-8}	$	\\
1218\tablenotemark{c}	&	9.5	$\pm$	0.7	&	1.07	$\pm$	0.06	&	140	$\pm$	14	&	[-0.8	$\pm$	0.1]	&	[-2.3	$\pm$	0.1]	&	$	4.45\times 10^{-7}  \pm  3.8\times 10^{-8}	$	\\
1303\tablenotemark{c}	&	21.2	$\pm$	2.1	&	0.60	$\pm$	0.06	&	472	$\pm$	172	&	-0.46	$\pm$	0.26	&	-1.5	$\pm$	0.1	&	$	4.27\times 10^{-7}  \pm  1.4\times 10^{-7}	$	\\
1683\tablenotemark{c}	&	3.9	$\pm$	0.5	&	5.90	$\pm$	0.10	&	325	$\pm$	[60]	&	-1.13	$\pm$	[0.1]	&	-2.4	$\pm$	[0.1]	&	$	3.05\times 10^{-6}  \pm  3.0\times 10^{-7}	$	\\
1734\tablenotemark{b}	&	46.7	$\pm$	2.8	&	1.09	$\pm$	0.06	&	94	$\pm$	6	&	-0.64	$\pm$	0.34	&	-3.1	$\pm$	0.5	&	$	2.98\times 10^{-7}  \pm  4.1\times 10^{-8}	$	\\
1819\tablenotemark{b}	&	53.0	$\pm$	0.6	&	1.73	$\pm$	0.06	&	728	$\pm$	73	&	[-0.8	$\pm$	0.1]	&	[-2.3	$\pm$	0.1]	&	$	1.30\times 10^{-6}  \pm  2.0\times 10^{-7}	$	\\
2035\tablenotemark{a}	&	5.1	$\pm$	0.8	&	0.59	$\pm$	0.05	&	107	$\pm$	11	&	[-0.8	$\pm$	0.1]	&	[-2.3	$\pm$	0.1]	&	$	1.75\times 10^{-7}  \pm  1.9\times 10^{-8}	$	\\
2047\tablenotemark{b}	&	41.1	$\pm$	2.1	&	1.36	$\pm$	0.06	&	123	$\pm$	13	&	-0.25	$\pm$	0.39	&	-2.8	$\pm$	0.4	&	$	3.86\times 10^{-7}  \pm  5.7\times 10^{-8}	$	\\
2080\tablenotemark{c}	&	53.8	$\pm$	0.7	&	4.08	$\pm$	0.09	&	337	$\pm$	27	&	-0.97	$\pm$	0.06	&	-3.1	$\pm$	0.8	&	$	1.92\times 10^{-6}  \pm  2.1\times 10^{-7}	$	\\
2203\tablenotemark{a}	&	15.1	$\pm$	3.4	&	1.13	$\pm$	0.06	&	194	$\pm$	47	&	-1.30	$\pm$	0.2	&	-2.6	$\pm$	0.9	&	$	4.43\times 10^{-7}  \pm  9.5\times 10^{-8}	$	\\
2345\tablenotemark{b}	&	89.0	$\pm$	5.2	&	1.84	$\pm$	0.06	&	150	$\pm$	13	&	0.18	$\pm$	0.28	&	[-2.3	$\pm$	0.1]	&	$	7.40\times 10^{-7}  \pm  9.0\times 10^{-8}	$	\\
2380\tablenotemark{a}	&	82.0	$\pm$	2.3	&	0.86	$\pm$	0.05	&	436	$\pm$	77	&	-0.18	$\pm$	0.24	&	[-2.0	$\pm$	0.1]	&	$	4.38\times 10^{-7}  \pm  1.2\times 10^{-7}	$	\\
2477\tablenotemark{c}	&	15.9	$\pm$	4.5	&	1.28	$\pm$	0.06	&	156	$\pm$	16	&	[-0.8	$\pm$	0.1]	&	[-2.3	$\pm$	0.1]	&	$	5.53\times 10^{-7}  \pm  4.5\times 10^{-8}	$	\\
2608\tablenotemark{a}	&	29.5	$\pm$	1.1	&	0.53	$\pm$	0.05	&	65	$\pm$	8	&	-0.50	$\pm$	[0.1]	&	-2.6	$\pm$	0.3	&	$	1.64\times 10^{-7}  \pm  2.4\times 10^{-8}	$	\\
2850\tablenotemark{a}	&	2.2	$\pm$	0.2	&	0.41	$\pm$	0.05	&	30	$\pm$	3	&	[-0.8	$\pm$	0.1]	&	[-2.3	$\pm$	0.1]	&	$	2.17\times 10^{-7}  \pm  3.0\times 10^{-8}	$	\\
2898\tablenotemark{a}	&	15.0	$\pm$	1.7	&	0.59	$\pm$	0.06	&	47	$\pm$	5	&	[-0.8	$\pm$	0.1]	&	[-2.3	$\pm$	0.1]	&	$	2.59\times 10^{-7}  \pm  3.3\times 10^{-8}	$	\\
2944\tablenotemark{a}	&	5.2	$\pm$	0.2	&	1.32	$\pm$	0.07	&	414	$\pm$	41	&	[-0.8	$\pm$	0.1]	&	[-2.3	$\pm$	0.1]	&	$	5.92\times 10^{-7}  \pm  7.3\times 10^{-8}	$	\\
3015\tablenotemark{c}	&	26.8	$\pm$	1.2	&	1.33	$\pm$	0.07	&	227	$\pm$	34	&	-0.73	$\pm$	0.21	&	-3.0	$\pm$	1.2	&	$	4.66\times 10^{-7}  \pm  1.2\times 10^{-7}	$	\\
3120\tablenotemark{a}	&	18.8	$\pm$	2.3	&	0.98	$\pm$	0.07	&	104	$\pm$	13	&	0.28	$\pm$	0.66	&	-2.5	$\pm$	0.3	&	$	2.66\times 10^{-7}  \pm  4.4\times 10^{-8}	$	\\
3142\tablenotemark{b}	&	33.1	$\pm$	4.1	&	1.16	$\pm$	0.06	&	441	$\pm$	[80]	&	-1.16	$\pm$	[0.1]	&	-1.9	$\pm$	[0.1]	&	$	4.15\times 10^{-7}  \pm  5.0\times 10^{-8}	$	\\
3257\tablenotemark{c}	&	63.6	$\pm$	0.8	&	2.92	$\pm$	0.07	&	256	$\pm$	26	&	[-0.8	$\pm$	0.1]	&	[-2.3	$\pm$	0.1]	&	$	1.55\times 10^{-6}  \pm  1.3\times 10^{-7}	$	\\
3283\tablenotemark{b}	&	45.8	$\pm$	1.4	&	1.98	$\pm$	0.05	&	204	$\pm$	27	&	-0.54	$\pm$	0.22	&	-2.5	$\pm$	0.4	&	$	7.24\times 10^{-7}  \pm  1.4\times 10^{-7}	$	\\
3405\tablenotemark{b}	&	67.4	$\pm$	19.2	&	1.17	$\pm$	0.05	&	511	$\pm$	139	&	-0.29	$\pm$	0.24	&	-1.8	$\pm$	0.2	&	$	1.00\times 10^{-6}  \pm  3.4\times 10^{-7}	$	\\
3439\tablenotemark{b}	&	150.8	$\pm$	1.7	&	1.60	$\pm$	0.05	&	133	$\pm$	13	&	[-0.8	$\pm$	0.1]	&	[-2.3	$\pm$	0.1]	&	$	6.72\times 10^{-7}  \pm  4.8\times 10^{-8}	$	\\
3663\tablenotemark{a}	&	204.4	$\pm$	0.3	&	3.05	$\pm$	0.06	&	245	$\pm$	23	&	-0.86	$\pm$	0.09	&	-2.3	$\pm$	0.2	&	$	1.43\times 10^{-6}  \pm  1.5\times 10^{-7}	$	\\
3853\tablenotemark{b}	&	91.3	$\pm$	14	&	2.09	$\pm$	0.08	&	599	$\pm$	187	&	-0.41	$\pm$	0.23	&	[-2.3	$\pm$	0.7	&	$	1.87\times 10^{-6}  \pm  8.8\times 10^{-7}	$	\\
5433\tablenotemark{b}	&	76.0	$\pm$	2.2	&	2.68	$\pm$	0.06	&	149	$\pm$	12	&	-0.40	$\pm$	0.17	&	[-2.0	$\pm$	0.1]	&	$	1.50\times 10^{-6}  \pm  1.7\times 10^{-7}	$	\\
5539\tablenotemark{b}	&	77.8	$\pm$	26.7	&	1.40	$\pm$	0.06	&	129	$\pm$	[30]	&	-0.64	$\pm$	[0.1]	&	-2.8	$\pm$	[0.1]	&	$	4.33\times 10^{-7}  \pm  3.1\times 10^{-8}	$	\\
5572\tablenotemark{a}	&	20.1	$\pm$	0.8	&	1.72	$\pm$	0.05	&	178	$\pm$	109	&	-1.03	$\pm$	0.74	&	[-2.0	$\pm$	0.1]	&	$	5.56\times 10^{-6}  \pm  9.2\times 10^{-8}	$	\\
6241\tablenotemark{a}	&	8.8	$\pm$	3.1	&	1.37	$\pm$	0.06	&	143	$\pm$	14	&	[-0.8	$\pm$	0.1]	&	[-2.3	$\pm$	0.1]	&	$	4.43\times 10^{-7}  \pm  3.5\times 10^{-8}	$	\\
6528\tablenotemark{a}	&	15.7	$\pm$	3.6	&	3.72	$\pm$	0.07	&	194	$\pm$	19	&	[-0.8	$\pm$	0.1]	&	[-2.3	$\pm$	0.1]	&	$	1.50\times 10^{-6}  \pm  1.5\times 10^{-7}	$	\\
7240\tablenotemark{c}	&	3.1	$\pm$	0.1	&	5.30	$\pm$	0.08	&	666	$\pm$	67	&	[-0.8	$\pm$	0.1]	&	[-2.3	$\pm$	0.1]	&	$	4.67\times 10^{-5}  \pm  6.7\times 10^{-6}	$	\\
7390\tablenotemark{c}	&	76.4	$\pm$	7.9	&	1.84	$\pm$	0.06	&	248	$\pm$	25	&	[-0.8	$\pm$	0.1]	&	[-2.3	$\pm$	0.1]	&	$	8.84\times 10^{-7}  \pm  7.7\times 10^{-8}	$	\\
8116\tablenotemark{c}	&	50.0	$\pm$	0.8	&	2.89	$\pm$	0.06	&	267	$\pm$	27	&	[-0.8	$\pm$	0.1]	&	[-2.3	$\pm$	0.1]	&	$	1.34\times 10^{-6}  \pm  1.2\times 10^{-7}	$	\\
\enddata
    
\tablenotetext{a}{This burst was chosen for its very high reported redshift (Band et al. 2004) based on the lag-luminosity relation.}  
\tablenotetext{b}{This burst was chosen for its very high reported redshift (Fenimore \& Ramirez-Ruiz 2000) based on the variability-luminosity relation.}
\tablenotetext{c}{This burst was chosen for its very high reported redshift (Yonetoku et al. 2004) based on the $E_{peak}$-luminosity relation.}
    
\end{deluxetable}

\clearpage

\begin{deluxetable}{lllll}
\tabletypesize{\scriptsize}
\tablecaption{Burst Luminosity Indicators
\label{tbl2}}
\tablewidth{0pt}
\tablehead{
\colhead{BATSE Trigger}   &
\colhead{$\tau_{lag}$ (s)}   &
\colhead{$V$}  &
\colhead{$\tau_{RTmin}$ (s)}  &
\colhead{$N_{peak}$} 
}
\startdata
111	&  $	3.488  \pm  0.800	$  &  $	0.0001  \pm  0.0010	$  &  $	5.12	\pm	1.00	$  &  $	1  \pm  0	$	\\
493	&  $	0.060  \pm  0.060	$  &  $	0.0023  \pm  0.0013	$  &  $	0.61	\pm	0.06	$  &  $	1  \pm  0	$	\\
1145	&  $	0.016  \pm  0.056	$  &  $	0.0039  \pm  0.0011	$  &  $	0.45	\pm	0.04	$  &  $	1  \pm  0	$	\\
1218	&  $	0.064  \pm  0.084	$  &  $	0.0033  \pm  0.0011	$  &  $	3.84	\pm	0.70	$  &  $	1  \pm  0	$	\\
1303	&  $	0.056  \pm  0.032	$  &  $	0.0265  \pm  0.0014	$  &  $	0.11	\pm	0.02	$  &  $	8  \pm  2	$	\\
1683	&  $	0.001  \pm  0.006	$  &  $	0.0131  \pm  0.0005	$  &  $	0.12	\pm	0.02	$  &  $	4  \pm  1	$	\\
1734	&  $	0.032  \pm  0.020	$  &  $	0.0296  \pm  0.0020	$  &  $	0.09	\pm	0.02	$  &  $	6  \pm  1	$	\\
1819	&  $	0.028  \pm  0.052	$  &  $	0.0127  \pm  0.0020	$  &  $	0.13	\pm	0.03	$  &  $	3  \pm  1	$	\\
2035	&  $	0.004  \pm  0.008	$  &  $	0.0330  \pm  0.0030	$  &  $	0.03	\pm	0.02	$  &  $	3  \pm  0	$	\\
2047	&  $	0.072  \pm  0.022	$  &  $	0.0193  \pm  0.0008	$  &  $	0.10	\pm	0.03	$  &  $	5  \pm  2	$	\\
2080	&  $	0.030  \pm  0.004	$  &  $	0.0133  \pm  0.0002	$  &  $	0.26	\pm	0.02	$  &  $	9  \pm  3	$	\\
2203	&  $	0.004  \pm  0.056	$  &  $	0.0100  \pm  0.0013	$  &  $	0.11	\pm	0.02	$  &  $	2  \pm  1	$	\\
2345	&  $	0.702  \pm  0.126	$  &  $	0.0201  \pm  0.0015	$  &  $	0.13	\pm	0.02	$  &  $	6  \pm  2	$	\\
2380	&  $	0.004  \pm  0.048	$  &  $	0.0165  \pm  0.0009	$  &  $	0.10	\pm	0.03	$  &  $	18  \pm  4	$	\\
2477	&  $	0.916  \pm  0.168	$  &  $	0.0047  \pm  0.0010	$  &  $	1.79	\pm	0.20	$  &  $	1  \pm  0	$	\\
2608	&  $	0.004  \pm  0.060	$  &  $	0.0121  \pm  0.0015	$  &  $	0.15	\pm	0.04	$  &  $	3  \pm  1	$	\\
2850	&  $	0.008  \pm  0.084	$  &  $	0.0049  \pm  0.0026	$  &  $	0.26	\pm	0.04	$  &  $	1  \pm  0	$	\\
2898	&  $	0.004  \pm  0.104	$  &  $	0.0127  \pm  0.0021	$  &  $	0.38	\pm	0.08	$  &  $	2  \pm  1	$	\\
2944	&  $	0.004  \pm  0.080	$  &  $	0.0244  \pm  0.0029	$  &  $	0.13	\pm	0.02	$  &  $	5  \pm  2	$	\\
3015	&  $	0.100  \pm  0.032	$  &  $	0.0262  \pm  0.0013	$  &  $	0.06	\pm	0.02	$  &  $	14  \pm  3	$	\\
3120	&  $	0.004  \pm  0.064	$  &  $	0.0399  \pm  0.0026	$  &  $	0.26	\pm	0.09	$  &  $	5  \pm  1	$	\\
3142	&  $	0.001  \pm  0.008	$  &  $	0.0203  \pm  0.0008	$  &  $	0.06	\pm	0.02	$  &  $	8  \pm  2	$	\\
3257	&  $	1.016  \pm  0.198	$  &  $	0.0093  \pm  0.0004	$  &  $	1.15	\pm	0.05	$  &  $	1  \pm  0	$	\\
3283	&  $	0.056  \pm  0.026	$  &  $	0.0384  \pm  0.0015	$  &  $	0.07	\pm	0.02	$  &  $	4  \pm  1	$	\\
3405	&  $	0.080  \pm  0.048	$  &  $	0.0171  \pm  0.0010	$  &  $	0.05	\pm	0.02	$  &  $	10  \pm  3	$	\\
3439	&  $	1.568  \pm  0.118	$  &  $	0.0068  \pm  0.0002	$  &  $	0.96	\pm	0.32	$  &  $	1  \pm  0	$	\\
3663	&  $	0.002  \pm  0.022	$  &  $	0.0082  \pm  0.0001	$  &  $	0.15	\pm	0.01	$  &  $	5  \pm  1	$	\\
3853	&  $	0.048  \pm  0.036	$  &  $	0.0239  \pm  0.0044	$  &  $	0.09	\pm	0.03	$  &  $	2  \pm  0	$	\\
5433	&  $	0.086  \pm  0.036	$  &  $	0.0029  \pm  0.0005	$  &  $	1.28	\pm	0.20	$  &  $	1  \pm  0	$	\\
5539	&  $	0.001  \pm  0.088	$  &  $	0.0062  \pm  0.0003	$  &  $	0.26	\pm	0.02	$  &  $	3  \pm  1	$	\\
5572	&  $	0.002  \pm  0.004	$  &  $	0.0125  \pm  0.0006	$  &  $	0.04	\pm	0.02	$  &  $	2  \pm  1	$	\\
6241	&  $	0.004  \pm  0.024	$  &  $	0.0253  \pm  0.0026	$  &  $	0.07	\pm	0.02	$  &  $	7  \pm  1	$	\\
6528	&  $	0.002  \pm  0.016	$  &  $	0.0246  \pm  0.0008	$  &  $	0.06	\pm	0.02	$  &  $	6  \pm  1	$	\\
7240	&  $	0.028  \pm  0.006	$  &  $	0.0077  \pm  0.0004	$  &  $	0.05	\pm	0.02	$  &  $	2  \pm  0	$	\\
7390	&  $	0.096  \pm  0.100	$  &  $	0.0230  \pm  0.0007	$  &  $	1.28	\pm	0.30	$  &  $	3  \pm  1	$	\\
8116	&  $	0.032  \pm  0.010	$  &  $	0.0225  \pm  0.0004	$  &  $	0.13	\pm	0.02	$  &  $	11  \pm  2	$	\\
\enddata
    
\end{deluxetable}

\clearpage

\begin{deluxetable}{lll}
\tabletypesize{\scriptsize}
\tablecaption{Burst Luminosity Relations
\label{tbl3}}
\tablewidth{0pt}
\tablehead{
\colhead{Relation}   &
\colhead{Formula}   &
\colhead{$\sigma_{LogL}$} 
}
\startdata
$\tau_{lag}$-L	&	$\log L = 52.26 - 1.01 \log[\tau_{lag}(1+z)^{-1}/0.1s]$	&	0.39 \\
$V$-L	&	$\log L = 52.49 + 1.77 \log[V(1+z)/0.02]$	&	0.40	\\
$E_{peak}$-L	&	$\log L = 52.21 + 1.68 \log[E_{peak} (1+z)300keV]$	&	0.36	\\
$\tau_{RTmin}$-L	&	$\log L = 52.54 - 1.21 \log[\tau_{RTmin}(1+z)^{-1}/0.1s]$	&	0.47	\\
$N_{peak}$-L	&	$\log L \ge 50.32 + 2 \log[N_{peak}]$ for $N_{peak} \ge 2$	&	 0	\\
\enddata
    
\end{deluxetable}

\clearpage

\begin{deluxetable}{llllllll}
\tabletypesize{\scriptsize}
\tablecaption{Derived Redshifts and the One-Sigma Confidence Intervals
\label{tbl4}}
\tablewidth{0pt}
\tablehead{
\colhead{Trigger}   &
\colhead{$z_{lag}$}   &
\colhead{$z_{V}$}   &
\colhead{$z_{Epeak}$}   &
\colhead{$z_{RT}$}   &
\colhead{$z_{Npeak}$}   &
\colhead{$z_{best}$}   &
\colhead{$z_{1\sigma low}$}
}
\startdata
111	&	1.1	(0.7-1.7)	&	0.1	(0.1-20)	&	5.5	(2.3-20)	&	0.9	(0.5-1.6)	&	$>$	0.0	&	1.4	(1.0-2.0)	&	1.0	\\
493	&	8.7	(	3.3	-	20	)	&	1.1	(	0.5	-	4.0	)	&	4.3	(	1.9	-	18.5	)	&	2.4	(	1.3	-	5.4	)	&	$>$	0.0	&	3.7	(	2.3	-	6.1	)	&	2.3	\\
1145	&	20	(	1.8	-	20	)	&	2.2	(	0.9	-	8.0	)	&	2.7	(	1.4	-	6.9	)	&	3.1	(	1.6	-	6.8	)	&	$>$	0.0	&	3.0	(	1.9	-	5.0	)	&	1.9	\\
1218	&	7.1	(	2.4	-	20	)	&	1.5	(	0.6	-	4.9	)	&	4.2	(	1.8	-	18.7	)	&	0.6	(	0.3	-	1.1	)	&	$>$	0.0	&	1.8	(	1.2	-	2.7	)	&	1.2	\\
1303	&	6.7	(	2.5	-	20	)	&	8.5	(	3.4	-	20	)	&	13.8	(	3.4	-	20	)	&	11.6	(	2.7	-	20	)	&	$>$	1.5	&	8.5	(	5.2	-	20	)	&	5.2	\\
1683	&	20	(	0.7	-	20	)	&	2.1	(	0.9	-	8.2	)	&	2.5	(	1.2	-	7.4	)	&	2.2	(	1.1	-	4.9	)	&	$>$	0.5	&	2.4	(	1.5	-	4.0	)	&	1.5	\\
1734	&	14.4	(	6.3	-	20	)	&	20	(	10.9	-	20	)	&	2.9	(	1.4	-	7.5	)	&	16.7	(	6.2	-	20	)	&	$>$	1.8	&	14.5	(	8.2	-	20	)	&	8.2	\\
1819	&	5.3	(	1.3	-	20	)	&	4.0	(	1.3	-	20	)	&	15.0	(	5.0	-	20	)	&	3.1	(	1.5	-	9.6	)	&	$>$	0.5	&	6.4	(	3.4	-	19.1	)	&	3.4	\\
2035	&	20	(	8.8	-	20	)	&	20	(	10.9	-	20	)	&	5.1	(	2.1	-	20	)	&	20	(	11.7	-	20	)	&	$>$	1.2	&	20	(	14.9	-	20	)	&	14.9	\\
2047	&	6.5	(	3.4	-	13.5	)	&	20	(	7.3	-	20	)	&	3.5	(	1.6	-	10.7	)	&	11.5	(	4.5	-	20	)	&	$>$	1.4	&	8.2	(	5.1	-	14.0	)	&	5.1	\\
2080	&	4.1	(	2.3	-	7.8	)	&	3.3	(	1.3	-	20	)	&	4.0	(	1.8	-	18.2	)	&	1.6	(	0.9	-	3.2	)	&	$>$	1.2	&	3.1	(	2.2	-	4.6	)	&	2.2	\\
2203	&	20	(	0.02	-	20	)	&	10.8	(	2.5	-	20	)	&	6.8	(	2.4	-	20	)	&	9.5	(	3.8	-	20	)	&	$>$	0.6	&	8.9	(	4.5	-	20	)	&	4.5	\\
2345	&	1.1	(	0.7	-	1.8	)	&	20	(	5.0	-	20	)	&	2.8	(	1.3	-	8.6	)	&	5.8	(	2.5	-	18.5	)	&	$>$	1.3	&	2.6	(	1.8	-	3.8	)	&	1.8	\\
2380	&	20	(	0.03	-	20	)	&	14.5	(	3.0	-	20	)	&	15.8	(	4.5	-	20	)	&	8.3	(	2.7	-	20	)	&	$>$	3.1	&	13.8	(	5.4	-	20	)	&	5.4	\\
2477	&	1.1	(	0.7	-	1.9	)	&	2.0	(	0.9	-	8.0	)	&	4.1	(	1.8	-	18.5	)	&	0.9	(	0.5	-	1.6	)	&	$>$	0.6	&	1.5	(	1.1	-	2.0	)	&	1.1	\\
2608	&	20	(	0.02	-	20	)	&	20	(	7.4	-	20	)	&	2.8	(	1.4	-	7.7	)	&	16.8	(	6.1	-	20	)	&	$>$	1.3	&	11.5	(	5.6	-	20	)	&	5.6	\\
2850	&	20	(	0.1	-	20	)	&	5.4	(	1.4	-	20	)	&	0.9	(	0.5	-	1.7	)	&	7.5	(	3.2	-	20	)	&	$>$	0.0	&	2.8	(	1.7	-	4.9	)	&	1.7	\\
2898	&	20	(	0.01	-	20	)	&	20	(	6.1	-	20	)	&	1.4	(	0.8	-	2.8	)	&	4.4	(	2.0	-	11.2	)	&	$>$	0.8	&	3.7	(	2.3	-	6.9	)	&	2.3	\\
2944	&	20	(	0.01	-	20	)	&	18.8	(	4.9	-	20	)	&	20	(	5.0	-	20	)	&	5.4	(	2.3	-	20	)	&	$>$	1.0	&	11.9	(	5.4	-	20	)	&	5.4	\\
3015	&	4.4	(	2.4	-	8.9	)	&	20	(	7.9	-	20	)	&	9.2	(	3.0	-	20	)	&	19.4	(	6.1	-	20	)	&	$>$	3.0	&	9.1	(	5.6	-	16.4	)	&	5.6	\\
3120	&	20	(	0.01	-	20	)	&	20	(	11.2	-	20	)	&	3.6	(	1.6	-	12.2	)	&	5.8	(	2.5	-	17.7	)	&	$>$	1.6	&	11.1	(	5.6	-	20	)	&	5.6	\\
3142	&	20	(	0.3	-	20	)	&	20	(	4.1	-	20	)	&	20	(	5.3	-	20	)	&	20	(	5.0	-	20	)	&	$>$	1.5	&	20	(	8.9	-	20	)	&	8.9	\\
3257	&	0.6	(	0.4	-	1.0	)	&	2.4	(	1.0	-	12.5	)	&	3.5	(	1.6	-	14.9	)	&	0.7	(	0.4	-	1.2	)	&	$>$	0.0	&	1.1	(	0.8	-	1.5	)	&	0.8	\\
3283	&	4.9	(	2.5	-	10.9	)	&	20	(	8.1	-	20	)	&	4.2	(	1.8	-	20	)	&	10.3	(	3.8	-	20	)	&	$>$	0.9	&	7.6	(	4.8	-	14.6	)	&	4.8	\\
3405	&	2.7	(	1.3	-	7.2	)	&	11.9	(	2.0	-	20	)	&	14.5	(	3.2	-	20	)	&	17.6	(	3.2	-	20	)	&	$>$	1.4	&	6.4	(	3.4	-	15.4	)	&	3.4	\\
3439	&	0.8	(	0.5	-	1.2	)	&	3.0	(	1.2	-	20	)	&	2.6	(	1.3	-	7.3	)	&	1.2	(	0.6	-	2.4	)	&	$>$	0.0	&	1.3	(	1.0	-	1.8	)	&	1.0	\\
3663	&	20	(	0.06	-	20	)	&	2.0	(	0.9	-	7.3	)	&	3.2	(	1.5	-	11.6	)	&	2.9	(	1.4	-	6.9	)	&	$>$	0.8	&	2.8	(	1.7	-	4.9	)	&	1.7	\\
3853	&	2.9	(	1.3	-	8.0	)	&	12.6	(	2.3	-	20	)	&	14.5	(	3.0	-	20	)	&	3.5	(	1.5	-	16.1	)	&	$>$	0.3	&	4.7	(	2.7	-	9.6	)	&	2.7	\\
5433	&	2.6	(	1.4	-	5.3	)	&	0.6	(	0.3	-	1.2	)	&	1.6	(	0.9	-	3.9	)	&	0.6	(	0.4	-	1.1	)	&	$>$	0.0	&	1.2	(	0.9	-	1.7	)	&	0.9	\\
5539	&	20	(	0.01	-	20	)	&	3.7	(	1.4	-	20	)	&	3.5	(	1.6	-	11.8	)	&	4.3	(	2.1	-	10.3	)	&	$>$	0.9	&	3.9	(	2.4	-	7.1	)	&	2.4	\\
5572	&	20	(	7.1	-	20	)	&	9.7	(	2.1	-	20	)	&	3.0	(	1.0	-	20	)	&	20	(	6.2	-	20	)	&	$>$	0.5	&	20	(	8.3	-	20	)	&	8.3	\\
6241	&	20	(	0.7	-	20	)	&	20	(	7.4	-	20	)	&	3.4	(	1.6	-	11.6	)	&	14.4	(	4.9	-	20	)	&	$>$	1.6	&	13.3	(	6.0	-	20	)	&	6.0	\\
6528	&	20	(	0.2	-	20	)	&	20	(	3.7	-	20	)	&	2.0	(	1.1	-	5.0	)	&	6.8	(	2.7	-	20	)	&	$>$	0.9	&	5.1	(	2.8	-	11.3	)	&	2.8	\\
7240	&	2.5	(	1.4	-	4.7	)	&	0.8	(	0.4	-	1.7	)	&	6.8	(	2.2	-	20	)	&	3.2	(	1.4	-	10.6	)	&	$>$	0.2	&	2.5	(	1.7	-	3.9	)	&	1.7	\\
7390	&	3.2	(	1.4	-	8.9	)	&	20	(	5.2	-	20	)	&	5.8	(	2.2	-	20	)	&	0.8	(	0.5	-	1.5	)	&	$>$	0.7	&	2.9	(	1.9	-	4.5	)	&	1.9	\\
8116	&	4.9	(	2.6	-	10.7	)	&	20	(	3.5	-	20	)	&	3.8	(	1.7	-	19.7	)	&	3.4	(	1.6	-	8.7	)	&	$>$	1.6	&	4.7	(	3.1	-	7.9	)	&	3.1	\\
\enddata
    
\end{deluxetable}

\end{document}